\documentclass[11pt]{article}
\usepackage{graphicx}

\newcommand{\BABARPubYear}    {00}

\newcommand{\BABARConfNumber} {09}
\newcommand{\SLACPubNumber} {8531}

\usepackage{relsize}
\def\babar{\mbox{\slshape B\kern-0.1em{\smaller A}\kern-0.1em
    B\kern-0.1em{\smaller A\kern-0.2em R}}}

\def\ellp       {\ensuremath{\ell^+}}

\def\ccbar {\ensuremath{c\overline c}}

\def\Kbar  {\kern 0.2em\overline{\kern -0.2em K}{}}

\def\Kzb   {\ensuremath{\Kbar^0}}
\def\KzKzb {\ensuremath{K^0 \kern -0.16em \Kzb}}
\def\Dz    {\ensuremath{D^0}}
\def\Dbar  {\kern 0.2em\overline{\kern -0.2em D}{}}

\def\Dzb   {\ensuremath{\Dbar^0}}
\def\DzDzb {\ensuremath{D^0 {\kern -0.16em \Dzb}}}
\def\Dstar   {\ensuremath{D^*}}

\def\Bz    {\ensuremath{B^0}}
\def\B     {\ensuremath{B}}
\def\Bbar  {\kern 0.18em\overline{\kern -0.18em B}{}}

\def\Bzb   {\ensuremath{\Bbar^0}}
\def\Bu    {\ensuremath{B^+}}
\def\Bub   {\ensuremath{B^-}}

\def\BB    {\ensuremath{B\Bbar}} 
\def\BzBzb {\ensuremath{B^0 {\kern -0.16em \Bzb}}}

\mathchardef\Upsilon="7107
\def\Y#1S{\ensuremath{\Upsilon{(#1S)}}}

\def\FourS {\Y4S}

\mathchardef\Deltares="7101
\mathchardef\Xi="7104
\mathchardef\Lambda="7103
\mathchardef\Sigma="7106
\mathchardef\Omega="710A
\def\Deltabar   {\kern 0.25em\overline{\kern -0.25em \Deltares}{}}
\def\Lbar {\kern 0.2em\overline{\kern -0.2em\Lambda\kern 0.05em}\kern-0.05em{}}
\def\Sigbar{\kern 0.2em\overline{\kern -0.2em \Sigma}{}}
\def\Xibar{\kern 0.2em\overline{\kern -0.2em \Xi}{}}
\def\Obar{\kern 0.2em\overline{\kern -0.2em \Omega}{}}
\def\Nbar{\kern 0.2em\overline{\kern -0.2em N}{}}
\def\Xbar{\kern 0.2em\overline{\kern -0.2em X}{}}

\def\ev   {\ensuremath{\rm \,e\kern -0.08em V}}
\def\kev  {\ensuremath{\rm \,ke\kern -0.08em V}} 
\def\mev  {\ensuremath{\rm \,Me\kern -0.08em V}} 
\def\gev  {\ensuremath{\rm \,Ge\kern -0.08em V}} 
\def\gevc {\ensuremath{{\rm \,Ge\kern -0.08em V\!/}c}} 
\def\tev  {\ensuremath{\rm \,Te\kern -0.08em V}}
\def\mevc {\ensuremath{{\rm \,Me\kern -0.08em V\!/}c}} 
\def\gevcc{\ensuremath{{\rm \,Ge\kern -0.08em V\!/}c^2}} 
\def\mevcc{\ensuremath{{\rm \,Me\kern -0.08em V\!/}c^2}}

\def\mum  {\ensuremath{\,\mu\rm m}} 

\def\invfb   {\ensuremath{\mbox{\,fb}^{-1}}}
\def\mus  {\ensuremath{\rm \,\mus}}

\def\ps   {\ensuremath{\rm \,ps}}

\def\mus        {\ensuremath{\,\mu{\rm s}}}    
\def\ps         {\ensuremath{{\rm \,ps}}}   
\def\degrees{\ensuremath{^{\circ}}}

\def\gsim{{~\raise.15em\hbox{$>$}\kern-.85em
          \lower.35em\hbox{$\sim$}~}}
\def\lsim{{~\raise.15em\hbox{$<$}\kern-.85em
          \lower.35em\hbox{$\sim$}~}}

\def\CP                 {\ensuremath{C\!P}}
\def\ra                 {\ensuremath{\rightarrow}}

\def\pep2{PEP-II}

\newcommand{\eqref}[1]{Eq.~(\ref{eq:#1})}

\newcommand{\pl}        [1]  {{Phys.\ Lett.\ {\bf #1}}}      

\newcommand{\prl}       [1]  {{Phys.\ Rev.\ Lett.\ {\bf #1}}} 

\newcommand{\zp}        [1]  {{Z.\ Phys.\ {\bf #1}}}

\def\jetset74   {\mbox{\tt Jetset \hspace{-0.5em}7.\hspace{-0.2em}4}}

\long\def\inst#1{\par\nobreak\kern 4pt\nobreak
    {\it #1}\par\vskip 10pt plus 3pt minus 3pt}

\def\Journal#1&#2&#3(#4){\unskip, #1~{\bf #2} (#4) #3}
 
\newcommand{\bi}{\begin{itemize}}
\newcommand{\ei}{\end{itemize}}
\newcommand{\ben}{\begin{enumerate}}
\newcommand{\een}{\end{enumerate}}
\newcommand{\bc}{\begin{center}}
\newcommand{\ec}{\end{center}}
\newcommand{\bt}{\begin{table}}
\newcommand{\et}{\end{table}}
\newcommand{\be}{\begin{equation}}
\newcommand{\eeq}{\end{equation}}
\newcommand{\ba}{\begin{eqnarray}}
\newcommand{\ea}{\end{eqnarray}}

\newcommand{\rar}{\ifmmode {\rightarrow} \else {$\rightarrow$}\fi}
\newcommand{\lar}{\ifmmode {\leftarrow} \else {$\leftarrow$}\fi}
\newcommand{\Ra}{\ifmmode {\Rightarrow} \else {$\Rightarrow$}\fi}
\newcommand{\La}{\ifmmode {\Leftarrow} \else {$\Leftarrow$}\fi}
\newcommand{\Lra}{\ifmmode {\Longrightarrow} \else {$\Longrightarrow$}\fi}
\newcommand{\Lla}{\ifmmode {\Longleftarrow} \else {$\Longleftarrow$\fi}}
\newcommand{\Llra}{\ifmmode {\Longleftrightarrow} \else {$\Longleftrightarrow$\fi}}
\newcommand{\Lk}{\ifmmode {{\cal L}} \else {${\cal L}$}\fi}
\newcommand{\Wt}{\ifmmode {{\cal W}} \else {${\cal W}$}\fi}
\newcommand{\Br}{\ifmmode {{\cal B}} \else {${\cal B}$}\fi}
\newcommand{\N}{\ifmmode {{\cal N}} \else {${\cal N}$}\fi}
\newcommand{\G}{\ifmmode {{\cal G}} \else {${\cal G}$}\fi}
\newcommand{\E}{\ifmmode {{\cal E}} \else {${\cal E}$}\fi}

\newcommand{\rhob}{\ifmmode {\bar\rho} \else {$\bar\rho$}\fi}
\newcommand{\etab}{\ifmmode {\bar\eta} \else {$\bar\eta$}\fi}
\newcommand{\dmd}{\ifmmode {\Delta m_d} \else {$\Delta m_d$}\fi}
\newcommand{\mnusq} {\ifmmode{{M_\nu}^2} \else {$M_{\nu}^2$}\fi} 
\newcommand{\BzBz} {\ifmmode {\Bz\bar{B^0_d}} \else { $\Bz{\bar{B^0_d}}$ }\fi }
\newcommand{\Bp} {\ifmmode {B^+} \else { $B^+$ }\fi }
\newcommand{\Bm} {\ifmmode {B^-} \else { $B^-$ }\fi }
\newcommand{\DelZ}{\ifmmode {\Delta Z} \else {$\Delta Z$}\fi}
\newcommand{\DelT}{\ifmmode {\Delta t} \else {$\Delta t$}\fi}
\newcommand{\DTau}{\ifmmode {\Delta \tau} \else {$\Delta \tau$}\fi}
\newcommand{\Zet}{\ifmmode {Z} \else {$Z$}\fi}
\newcommand{\X}{\ifmmode {X} \else {$X$}\fi}
\newcommand{\Vcb}{\ifmmode {|V_{cb}|} \else {$|V_{cb}|$}\fi}
\newcommand{\Dsp} {\ifmmode {D^{*-}} \else { $D^{*-}$ }  \fi}
\newcommand{\pstar} {\ifmmode {\pi^{*}} \else { $\pi^{*}$ } \fi }
\newcommand{\tBz}{\ifmmode {\tau_{\B^0}} \else { $\tau_{\B^0}$ } \fi }
\newcommand{\tBp}{\ifmmode {\tau_\Bp} \else { $\tau_\Bp$ } \fi }
\newcommand{\Dsst}{\ifmmode {D^{**}} \else { $D^{**}$ } \fi }

\newcommand{\BtoDs}{\mbox{$B^0\rightarrow D^{*-} \ell^+\nu_\ell$}}
\newcommand{\DsptoDz}{\mbox{$D^{*-}\rightarrow D^{0} \pi^-$}}
\newcommand{\BtoDss}{\mbox{$B\rightarrow D^{**} \ell^+ \nu_\ell$}}
\newcommand{\bb}{\ifmmode {B\bar{B}} \else { $B\bar{B}$ } \fi }
\newcommand{\plab}{\ifmmode{p} \else {$p$} \fi}
\newcommand{\ks}{\ifmmode{k^*} \else {$k^*$} \fi}
\newcommand{\micron}{\ifmmode{\mu m} \else {$\mu m$} \fi}
\newcommand{\Ecal}{\ifmmode{E_{cal}} \else {$E_{cal}$} \fi}
\newcommand{\yfs}{\ifmmode{\Upsilon(4S)} \else {$\Upsilon(4S)$} \fi }
\newcommand{\GeVc}{\ifmmode{GeV^2/c^4} \else {$GeV^2/c^4$} \fi }
\newcommand{\BpBm}{\Bu\Bub}

\def\etal{{\it et al.}}
\def\BDstarpi{$\Bz\rightarrow D^{*-} \pi^{+}$}
\def\DstarDpi{$D^{*-} \rightarrow \Dzb\pi^{-}$}
\def\Dknpi{$\Dzb\rightarrow K^-\pi^{+},   K^- \pi^{+}\pi^{0}, 
  K^- \pi^{-}\pi^{+}\pi^{+},  \overline K^0 \pi^{-}\pi^{+}, ...$}
\def\Br{{\cal B}}

\begin{document}
{\pagestyle{plain}

\begin{flushright}
\babar-CONF-\BABARPubYear/\BABARConfNumber \\
SLAC-PUB-\SLACPubNumber
\end{flushright}

\par\vskip 3cm

\begin{center}
\Large \bf Measurement of the \boldmath $B^0$ meson properties using partially
reconstructed  $\Bz$ to $D^{*-}\, \pi^{+}$ 
and $\Bz$ to $D^{*-}\,\ell^+\,\nu_\ell$
decays \\
with the \babar\ detector 
\end{center}
\bigskip

\begin{center}
\large The \babar\ Collaboration\\
\mbox{ }\\
July 25, 2000
\end{center}
\bigskip \bigskip

\begin{center}
\large \bf Abstract
\end{center}

\noindent
The two $B^0$ decay processes \mbox{$B^0\rightarrow D^{*-}\pi^+$} 
and \mbox{$B^0\rightarrow D^{*-} \ell^+ \nu_{\ell}$} have been studied 
by means of a partial reconstruction technique 
using a data sample collected with the \babar\ detector at the \pep2\ 
storage ring. To increase statistics, only the soft $\pi^-$ from the
decay \mbox{$D^{*-} \rightarrow \pi^- D^0$} was used  
in association with either an oppositely-charged high-momentum
pion or lepton. Events were then identified by exploiting the constraints from the
simple kinematics of \FourS\ decays. A clear signature is obtained in each case.  The position of the 
$B^0$ decay
point was obtained from the reconstructed $\pi^+(\ell^+) \pi^-$
vertex. The position of the other \Bzb\ in the event was also determined.
Taking advantage of the boost given to the
\FourS\ system by the asymmetric beam energies of \pep2, the
 lifetime of the $B^0$ meson has been measured from the separation distance
 between the two vertices along the beam direction. 
The preliminary results are:
\begin{eqnarray}
\nonumber \tau_{B^0} &=& 1.55 \pm 0.05 \pm 0.07 ~\mathrm{ps}, \\
\nonumber \tau_{B^0} &=& 1.62 \pm 0.02 \pm 0.09 ~\mathrm{ps},
\end{eqnarray}
respectively for the \mbox{$B^0\rightarrow D^{*-}\pi^+$} and
 \mbox{$B^0\rightarrow D^{*-} \ell^+ \nu_{\ell}$} channels.\\

\vfill
\begin{center}
Submitted to the XXX$^{th}$ International 
Conference on High Energy Physics, Osaka, Japan.
\end{center}

\newpage
}

\begin{center}
\small

The \babar\ Collaboration
\bigskip

B.~Aubert,
A.~Boucham,
D.~Boutigny,
I.~De Bonis,
J.~Favier,
J.-M.~Gaillard,
F.~Galeazzi,
A.~Jeremie,
Y.~Karyotakis,
J.~P.~Lees,
P.~Robbe,
V.~Tisserand,
K.~Zachariadou
\inst{Lab de Phys.\ des Particules, F-74941 Annecy-le-Vieux, CEDEX, France}
% ...

\end{center}\newpage

\setcounter{footnote}{0}

%
%

\section{Introduction}

This document presents two analyses designed to select large samples of
$B^0$ mesons with inclusive reconstruction techniques. The first method 
finds \BDstarpi\ events and the second \BtoDs; while the two techniques
are different in detail, they both share the common feature of
making no attempt to reconstruct
the $D^0$ produced in the  \DsptoDz\ decay, thereby achieving high 
efficiency comparing to the exclusive reconstruction.

Taking advantage of the boost of the $\Upsilon(4S)$ system at \pep2, 
the lifetime
of the $B^0$ meson is determined separately, with a good statistical precision,
using both samples of events. 
Tagging the flavor of the second \Bz\ in the event 
will eventually also allow a measurement of the mixing parameter $\Delta m_d$.

\subsection{\boldmath \BDstarpi\ inclusive reconstruction}

The decay \BDstarpi\ is interesting for many reasons, but particularly since
it may exhibit \CP\ violation~\cite{physbook_dstarpi}.
Since the  asymmetry is expected to be small, this is a long term goal
and the first steps toward it are the measurement
of the $B^0_d$ lifetime and mixing using this process.

The full reconstruction 
of the decay chain \BDstarpi, \DstarDpi, where the $D^0$ decays in the mode  
\Dknpi,
allows one to evaluate detector performance
and to extract the signal with an excellent signal over background ratio.
However the loss in statistics is substantial and may limit
a \CP\ violation measurement given the small expected asymmetry.

An alternative approach exists for extracting the signal without 
reconstructing the decay of the \Dz\ meson. This partial reconstruction method
has already been used successfully by other experiments \cite{ARGUS:1,CLEO:1} 
and allows an increase of the size of the reconstructed sample
by about an order of magnitude while
maintaining the background at a reasonable level.
 
No attempt is made to reconstruct the 
\Dz\ decays. Therefore, one searches for a pair of oppositely-charged pions
($\pi_f$, $\pi_s$) and, assuming that their origin is a \Bz\ meson, 
calculates the missing invariant mass which should be the \Dz\ mass if
our hypothesis was correct. 
Without the constraint of the \Dz\ mass, the direction of
the \B\ meson is unknown. Although its angle with respect to $\pi_f$ 
direction can be deduced, 
the angle $\phi$ around this direction is undetermined.
For more details see 
\cite{ARGUS:1,CLEO:1}.
 
Using the beam energy constraint,
the missing mass, $M_{miss}$, 
is computed from the energy and momenta of the two
reconstructed pions. Since this still depends on the 
unknown angle $\phi$ of the
\Bz\ momentum, 
in this analysis, $M_{miss}$ is defined as the average of the maximum and 
minimum value of $M_{miss}$ over all $\phi$ angles. 

A Monte Carlo 
study shows that the resolution on $M_{miss}$ obtained in this way 
is of the order of 3\mevcc, and that it is dominated by the 
tracking precision for the measurement of the two pions.

\subsection{\boldmath \BtoDs\ inclusive reconstruction}

The \Bz\ semileptonic decay process \BtoDs\ is characterized by high branching ratio and clear
 experimental signature. These facts allow the selection of large event samples 
with small background 
contamination. Therefore this process has been widely used in the past to determine 
several \Bz\ properties, along which its lifetime, \tBz, and the \BzBzb\
mass splitting, \dmd~\cite{Blife}. In those measurements the \Dsp  particle was reconstructed 
through its decay \DsptoDz, and the \Dz\ was identified by means of  
exclusive reconstruction of a few final states which 
provide high selection efficiency and high background rejection. 

Due to the restricted phase space of the \DsptoDz\ transition it is however possible to tag this
process by reconstructing only the low energy charged pion (hereafter called \pstar) from the \Dsp  
decay, without reconstructing the \Dz. This ``inclusive'' approach allows us to select an event  
sample about ten times more abundant than the  exclusive analysis. This
method has been extensively used in the past by the ARGUS and CLEO collaborations at the \yfs\ to
determine several \Bz\ and \Dz\ properties \cite{ARGUS,CLEO}, and by DELPHI and OPAL at LEP to
measure \Vcb\ and the \BtoDs\ branching fraction \cite{DFVcb,OPVcb}. Measurements of the \Bz\ lifetime 
and oscillation frequency with this approach have only been performed by the DELPHI collaboration 
at LEP \cite{DFtau,DFosc} and are, at present time, among the most precise determinations of those
quantities.



\section{Detector and data}

The data used in this analysis were collected with the \babar\ detector 
(for a detailed description see \cite{TDR}) 
at the \pep2\ storage ring. The statistics analyzed corresponds to an 
integrated luminosity of 
7.9 (7.1) \invfb\ of data collected on the \FourS peak, and 1.2 
(1.0) \invfb\ collected below peak, 
for the  \BDstarpi\ (\BtoDs) analysis.
The off-resonance data were used for background study purposes. 

Samples of about one million \BzBzb\ and \BpBm\ simulated events were
also analyzed; in addition, about 28000 (175000) \BzBzb\ were generated, 
were one of the \Bz\ mesons was 
forced to decay to \BDstarpi\  ($\Dsp\ellp \nu$).


\section{\boldmath \Bz\ to $D^{*-}\pi^{+}$ analysis and results}

\subsection{Inclusive reconstruction of \boldmath \BDstarpi}
\label{sec:inclusiverec}

\subsubsection{Track and signal selection}

The fast and slow pion have been selected by requiring Drift Chamber
(DCH) 
and Silicon Vertex Detector (SVT) quality cuts, respectively.
In addition, the fast pion should have a momentum in the center of
mass frame between 2.114 and 2.404\gevc, and should 
not to be identified as a muon or as an electron.
The slow pion should have a momentum in the laboratory of at least 50\mevc.
A further set of cuts were applied to reject electrons 
and kaons from the fast and slow pion tracks using the calorimeter and
the Detection of Internally Reflected Cherenkov light (DIRC).

In order to obtain the signal, pairs of oppositely-charged tracks
were required and the inclusive reconstruction method was 
applied, leading to a value for $M_{miss}$.
An optimized selection criterium is necessary in order to reduce 
the combinatorial background, mainly from  continuum events.

In order of decreasing discriminating power, the quantities used for
signal selection are:
\begin{enumerate}
\item $R_2$, the normalized second 
Fox-Wolfram moment, is required to be less than 0.35.
\item Requiring that no other tracks be in a cone of opening angle 0.4 rad
  centered on the fast pion candidate momentum vector in the \FourS\ system.
\item Two sets of particles are defined: all tracks and calorimeter
  clusters,  excluding the fast and slow pion (SET1), and
 a set where all particles in a cone around the reference 
$D^0$ direction are also excluded (SET2) (see section~\ref{sec:conecut}).
A single discriminating variable is constructed
using the Fisher method, based on 15 input variables.
(For SET1, the scalar sum of the CM momenta of tracks and showers in nine
$20^0$ angular bins around the fast pion, and the sphericity. For SET2, the
angle of the sphericity axis and the angle of the particle of highest energy
with respect to the fast pion, and the combined mass and momentum of  all the
particles.)
\item The cosine of the helicity angle\footnote{angle between the soft pion 
and the $D^*$ flight direction in the rest frame of the latter} 
is required to be larger than 0.4 in absolute value.
\end{enumerate}
The efficiency for these cuts as determined using Monte Carlo simulation is 27.5\%.

\begin{figure}[hbt]
\begin{minipage}{.45\linewidth}
\begin{minipage}[h]{1.0\textwidth}
\begin{center}
\mbox{\includegraphics[height=7.cm]{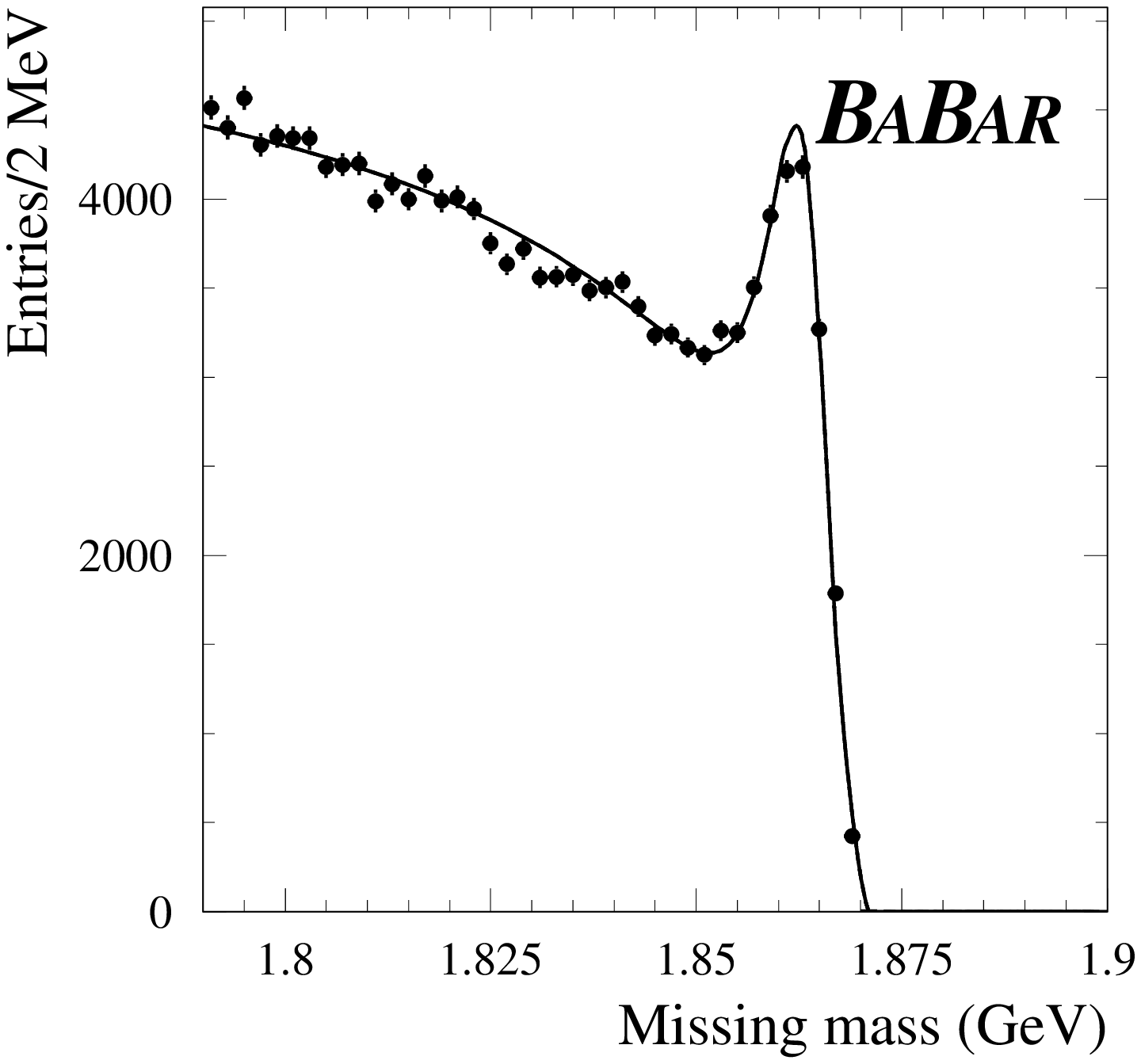}}
\end{center}
\vspace*{-1.cm}
\caption{Missing mass distribution
for partially reconstructed \BDstarpi\ events from data.
Here only the cut on $R_2$ is applied.}
\label{fig:bpidstar_mrec_1}
\end{minipage}
\end{minipage}
\hspace{0.5cm}
\begin{minipage}{.45\linewidth}
\vspace{-0.5cm}
\begin{minipage}[h]{1.0\textwidth}
\begin{center}
\mbox{\includegraphics[height=7.cm]{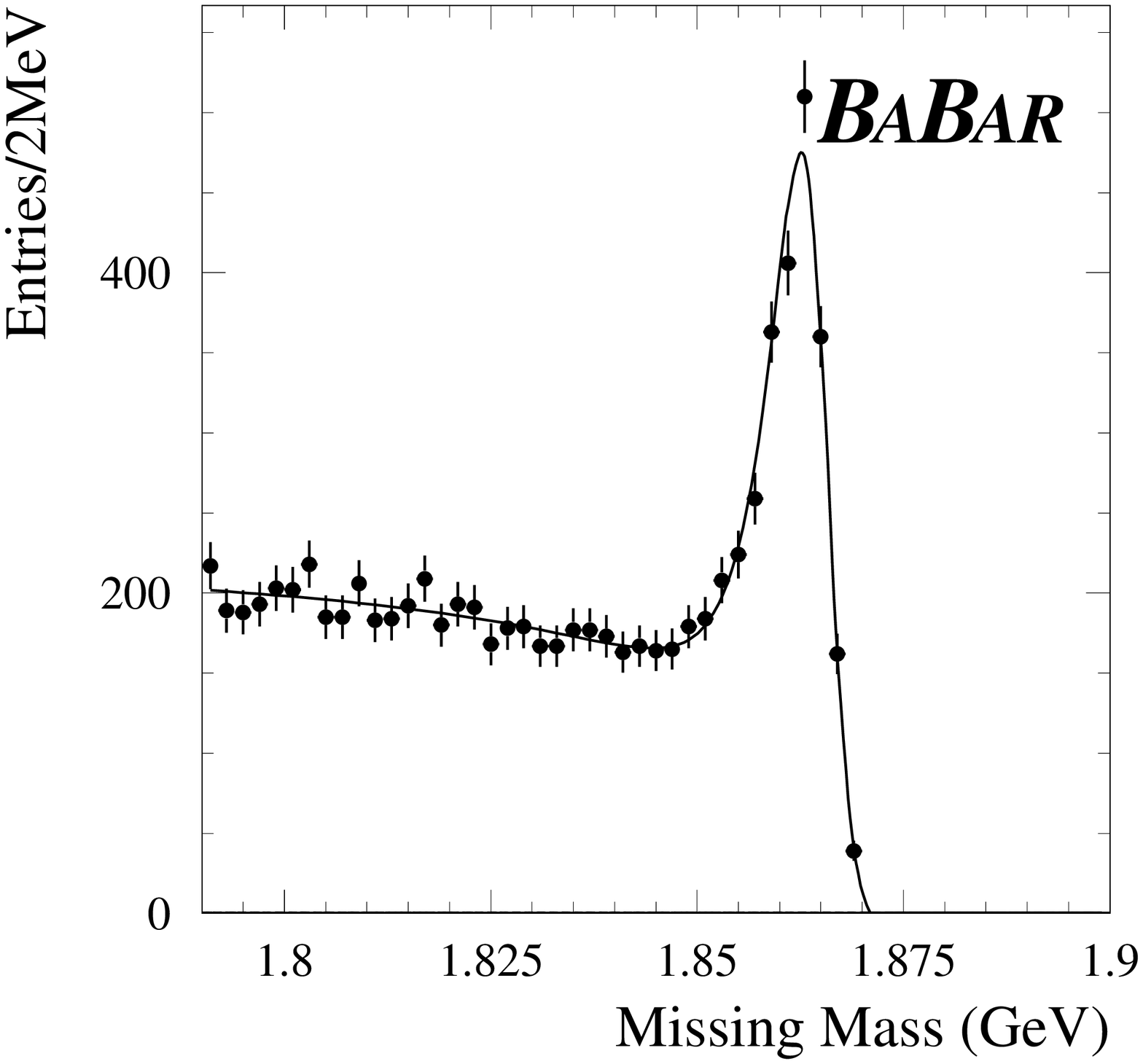}}
\end{center}
\vspace*{-1.cm}
\vfill
\caption{Missing mass distribution
for partially reconstructed  \BDstarpi\ events from data 
after all the cuts are applied. 
} 
\label{fig:bpidstar_mrec_2}
\end{minipage}
\end{minipage}
\vspace*{0.2cm}
\end{figure}

The distribution of the missing mass obtained on the data,
applying only the $R_2$ cut, is shown in Fig.~\ref{fig:bpidstar_mrec_1}. The fitted signal is
$10992 \pm 235$ events. Applying all the cuts (see also Section 6.4)
the distribution in Fig.~\ref{fig:bpidstar_mrec_2} is obtained, containing $1696 \pm 107 $ events.

No peak is observed in the corresponding distribution for wrong charge
combinations, nor in the off-resonance data.

\subsubsection{Signal and background composition \label{sec:compo} }

Other \B\ decay modes may produce an enhancement
at the end of the missing mass spectrum in the same region where the signal is expected.
An obvious example is the decay $B^0 \rightarrow D^{*-} \rho^+$.

To study this effect,
we have fit separately the distribution obtained for generic $B$ mesons 
and charged \B\ mesons from Monte Carlo simulation.
Contributions to the missing mass spectrum come from a) \BDstarpi\ signal,
b) combinations of a true soft pion from a $D$ with a pion coming 
from a $B^0$ meson or from a direct $\rho$,
c) combinations of a true soft pion with a direct lepton, and
d) all other combinations.
Only the categories a), b) and c) show clearly a peak at the end of the spectrum.
The fit gives $ 461 \pm 26 $ events, which is consistent with the sum 
of the contribution from a) (365 events) and b)  (100 events). 
Therefore, we estimate that the signal contribution
is enhanced by 26\% from other $B^0$ decay modes.

No clear peaking is observed for charged \B\ mesons. The estimate 
for their fraction in the signal sample is $6.0 \pm 3.1$\%. 


Figure~\ref{fig:total1} shows a comparison between
data and Monte Carlo distributions for missing mass,
after all cuts. The Monte Carlo is normalized to the data.

\begin{table} [hb]
  \centering
\caption{ Contribution (in \%) of different background components in the 
sideband, signal region and for the wrong-charge combination 
for the ($\pi_f$, $\pi_s$) pair.} \vspace{0.3cm}
  \begin{tabular}{|l|c|c|c|}
    \hline
    component &  sideband & signal &  wrong charge 
    \\
    \hline\hline
    $q \overline{q}$, $q=u$, $d$, $s$  & 40.9  &  17.9 &  40.8   \\ 
    $c \overline{c}$                   & 17.4  &  30.6 &  19.1   \\
    $B^+ B^-$  generic                 & 24.3  &  28.2 &  21.3 \\
    \BzBzb\ generic                    & 17.4  &  23.3 &  18.8 \\    
    \hline
  \end{tabular}
\label{table:compo} 
\end{table}

For the lifetime measurement, the combinatorial 
background parameters have been fitted to the data 
for wrong-charge combinations with $ M_{miss} > 1.84 $\gevcc.
As a cross-check, a sideband region 
($1.8 < M_{miss} < 1.84 $\gevcc) is also defined.
The signal region is taken to be the interval $M_{miss} > 1.854$\gevcc.
The fractions of the different components in
the sideband, signal and wrong charge region 
are shown in Table~\ref{table:compo}.


\subsection{Lifetime measurement}

\subsubsection{\boldmath \BDstarpi\ vertex determination}

The \BDstarpi\ vertex has been computed fitting the fast and slow pion
tracks with the beam spot constraint. To take
into account the flight of the \B\ meson in the $xy$ plane, the effective size of the luminous region 
in the vertical direction has been assumed to be
60\mum.

The resolution on the \BDstarpi\ vertex is dominated by the beam spot and by the fast 
pion track. Therefore, we have required the fast pion to have at least
3 SVT hits. The longitudinal or $z$ precision on this vertex is strongly
correlated to the direction of the fast pion track 
and is 
best when this direction lies in the transverse plane, in which case the resolution is 
50\mum.
The error on the $z$ coordinate of this vertex is required to 
be less than 150\mum.
The $z$ resolution of the  \BDstarpi\ vertex can be fitted with the sum
of two Gaussian distributions, where the core Gaussian has a width of 
48\mum\ and contains 80\% of the events.




\subsubsection{Rejection  of \boldmath $D^0$ tracks and $B_{tag}$ vertex determination \label{sec:conecut}}

Before fitting the vertex of the second B in the event,
hereafter named  $B_{tag}$, it is necessary to 
reject tracks coming from the non-reconstructed $D^0$.
Including these tracks in the fit would result in a bias,
pulling the position of the $B_{tag}$ vertex systematically closer
to the \BDstarpi\ vertex and thereby reducing the measured lifetime.
Thanks to the high momentum of the $D^0$, most of its decay
products will be boosted in a cone around the parent direction. 
To reject these, tracks are excluded that lie
within the region in the $\theta_{TK}-p$ plane 
that is populated by $D^0$ decay particles, where $\theta_{TK}$ is the
the angle between the direction of a given track 
and the reconstructed  $D^0$ direction and $p$ is the momentum
of the track in the center of mass frame. 
This requirement is applied up to a
maximum angle of 2~rad.
It has an efficiency of 45\% for the tracks from the $B_{tag}$
and 1.8\% for the $D^0$ decay products. As a result, 98\% of the accepted tracks 
come from the $B_{tag}$.


The $B_{tag}$ vertex has been fitted from the tracks selected as 
explained above, discarding the tracks with excessive contribution to
$\chi^2$. To achieve a high quality vertex it is required that 
the fit be performed using at least two tracks.
Even after this treatment,
some tracks from the decay of charmed hadrons are still included
in the $B_{tag}$ vertex fit. As these come from the charm
decay vertex, which is generally at larger $z$ values due to the 
charm hadron lifetime and the boost in the laboratory frame, 
this produces a bias in the mean value of the $z$ coordinate of the order of
20\mum.
The $z$ resolution for the $B_{tag}$ vertex can be fit with the sum
of two Gaussian distributions, where the core Gaussian has a width of 
92\mum\ and it contains 70\% of the events.




\subsubsection{Vertex quality selection\label{section:lifetimefit}}

Further selection cuts are applied to improve the vertex quality.
The number of SVT hits associated to the fast pion is required
to be at least 4, the $z$ error on the \BDstarpi\ vertex fit is
required to be less than 150\mum, the normalized $\chi^2$ of the
\BDstarpi\ vertex fit is required to be less than 20, and
at least two tracks are required to be used by the $B_{tag}$ vertex fit.
The efficiency for these cuts estimated from Monte Carlo simulation to be 58\%.
The number of signal events is $1696 \pm 107$ and the fraction of 
combinatorial background
in the signal region $\alpha_{back}$ is $27\pm 6$\%.

\subsubsection{\boldmath $B^0$ lifetime fit}

Using the fit result for the two \B\ vertices, it is possible to extract a 
longitudinal vertex separation, $\Delta z = 
z_{D^* \pi} - z_{tag}$.
$\Delta t$ is then computed using the approximation $\Delta t =
\Delta z / \beta \gamma c $,  where the boost $\beta \gamma$ of the 
\FourS\ at \pep2\ is known to be 0.56. 
In order to fit the $\Delta t $ distribution, three regions were defined 
in the $M_{miss}$ spectrum: a signal region for candidates with 
$M_{miss} > 1.854$\gevcc, a sideband region with
$1.800< M_{miss} < 1.840$\gevcc\ and a wrong charge region for 
candidates with both pions have the same charge and 
$M_{miss}> 1.840$\gevcc. 

The wrong-charge distribution is used to define the shape of the combinatorial
background under the signal. The $\Delta t $ distribution of the
wrong charge region is fit with the function:
\begin{eqnarray}
\lefteqn {
f_{b}(\Delta t) =  (1-\alpha_B) (\alpha_{uds}\,  gauss(\Delta t,\sigma_{uds}) +
(1-\alpha_{uds})\,  gauss(\Delta t,\sigma_{c}))
 }
\nonumber  \\
& &  + \alpha_B 
( (1-\alpha_{bw} ) E_G ( \Delta t- b_n ; \tau_{back}, \sigma_{bn}) 
  +  \alpha_{bw} E_G ( \Delta t-b_w; \tau_{back}, \sigma_{bw}) )
\end{eqnarray}
where  $E_G ( t ; \tau , \sigma )$ is the convolution of an exponential with
lifetime $\tau$ with a gaussian of width $\sigma$. 

The parameters $\alpha_{bw}$, $b_n$, $\sigma_{bn}$, $b_w$, $\sigma_{bw}$ 
for the background ``resolution
function'' are fixed to the resolution function determined from Monte Carlo
simulation for signal events. 
The background lifetime is fixed to the value, 1.55\ps, measured from the same
distribution for \BB\ events in the Monte Carlo sample.
The widths $\sigma_{uds}$ and $\sigma_{c} $ of the two Gaussian distributions 
used to parametrise the continuum light quark ($uds$) and charm
components are fixed to the values, 0.70 and 0.96\ps\ respectively, extracted 
from the Monte Carlo.
$\alpha_B$ and $\alpha_{uds}$ are the only parameters left free in the
fit of the background $\Delta t$ distribution.

To take into account the fact that the fractions of $uds$ and charm in the different
samples are different (cf. table~\ref{table:compo}), 
$\alpha_{uds}$ is fixed to the value
0.37, obtained from the Monte Carlo simulation
in the signal region, for the final lifetime fit. 

The lifetime fit is performed using an event-by-event resolution estimate 
obtained by scaling 
the error given by the vertex fit for the two vertices. The Monte Carlo
$\Delta z$ pull distribution has been used for the rescaling. This 
can be described by two Gaussian distributions with widths $ p_n =1.04$  and $p_w =2.51$, 
centered at
$ s_n =-0.22$  and $s_w =-0.95$ respectively. The fraction $\alpha_{wide} $
of the events in the
wide  Gaussian is  11\%.

Therefore, the probability density function for the unbinned maximum
likelihood fit for event $i$ is: 
\begin{equation}
f(\Delta t_i, \sigma_i) = (1-\alpha_{back}) f_s(\Delta t_i,\sigma_i) + \alpha_{back} 
f_b (\Delta t_i) 
\end{equation}
where
\begin{equation}
f_s(\Delta t_i,\sigma_i) = (1-\alpha_{wide}) E_G ( \Delta t_i - s_n \sigma_i ; 
\tau_{B^0} ,S p_n  \sigma_i ) + \alpha_{wide} E_G ( \Delta t_i - s_w \sigma_i ; 
\tau_{B^0} ,S p_w  \sigma_i ) 
\end{equation}
Here  $\Delta t_i$ and $ \sigma_i$ are $\Delta t$ and its error 
coming from the fit of the two \B\ vertices and S is a global scale factor for 
the errors. 

\begin{figure}[hbt]
\vspace{0.5cm}
\begin{minipage}{.45\linewidth}
\begin{minipage}[h]{1.0\textwidth}
\begin{center}
\mbox{\includegraphics[height=7cm]{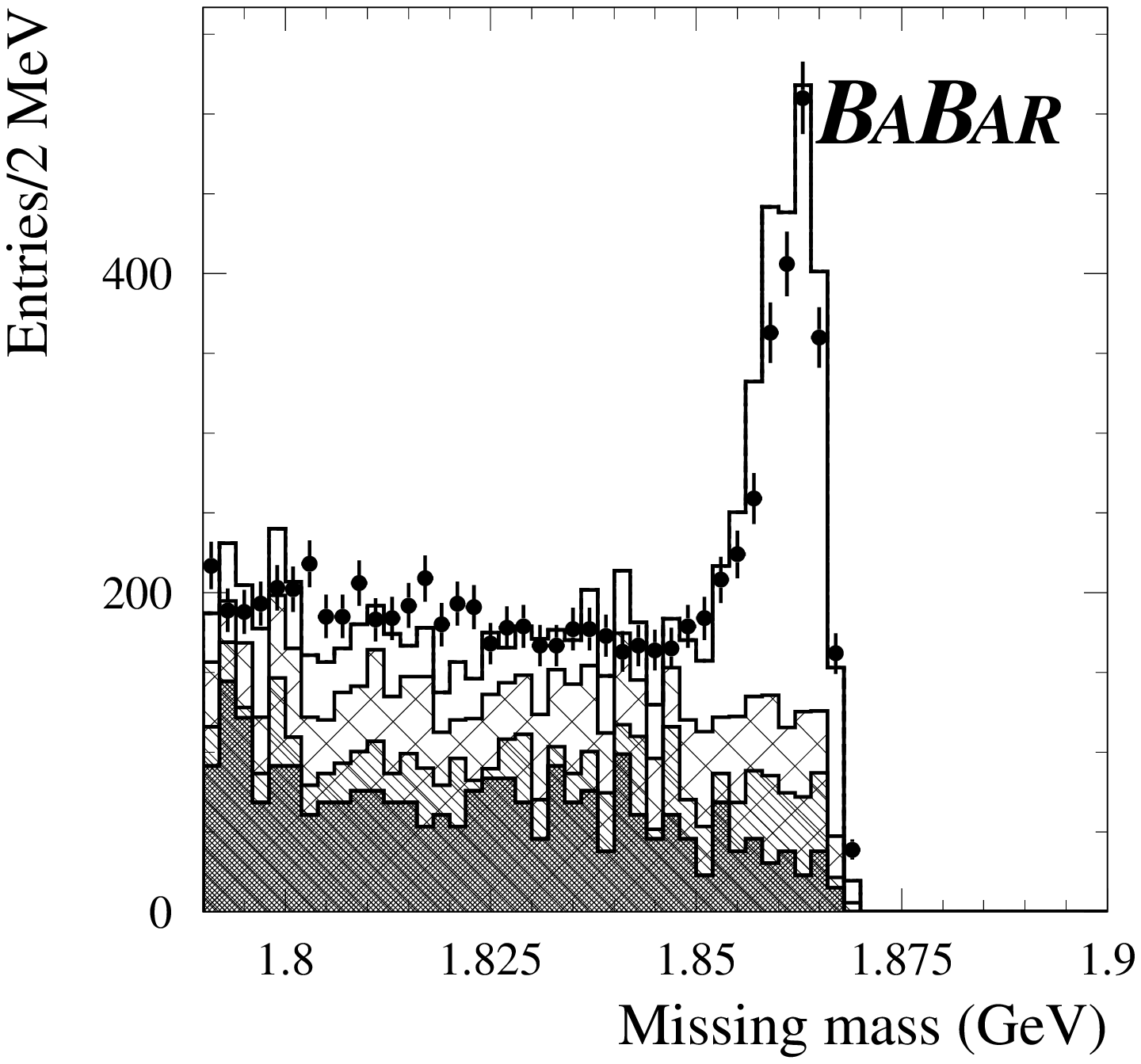}}
\end{center}
\vspace*{-0.8cm}
\caption{Missing mass distribution
  in partially reconstructed  \BDstarpi \ events after all cuts. 
  The data are represented by points.
The contribution of the $q \overline{q}$ final states from Monte Carlo is:
$uds$ (black),  $c$ (dense hatched), charged $B$ (light
  hatched), neutral $B$ (white).}
\label{fig:total1}
\end{minipage}
\end{minipage}
\hspace{0.5cm}
\begin{minipage}{.45\linewidth}
\begin{minipage}[h]{1.0\textwidth}
\vspace*{-0.3cm}
\begin{center}
\mbox{\includegraphics[height=7.cm]{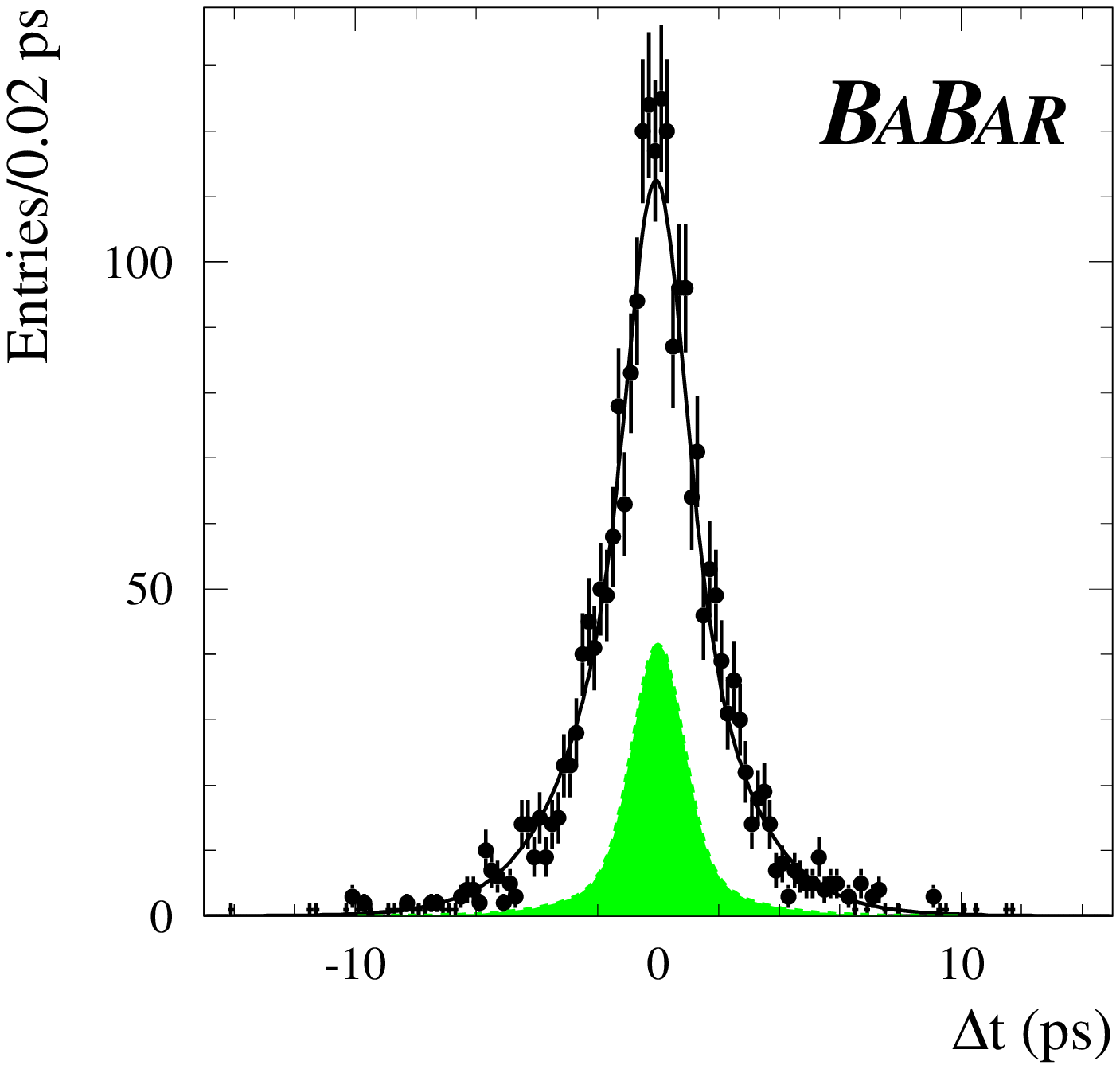}}
\end{center}
\vspace*{-0.8cm}
\caption{Distribution of $\Delta t$ in ps.  The data are represented by the points
with error bars.
The continuous line shows the result of the fit  and the shaded area 
shows the contribution of the combinatorial background (27\%
of the sample). 
} 
\label{fig:lifetime}
\vfill
\end{minipage}
\end{minipage}
\vfill
\end{figure}

The result of the unbinned maximum likelihood fit with one free parameter, the $B^0$ lifetime, is 
$\tau_{B^0} = 1.51 \pm 0.05$\ps\ as shown in Fig~\ref{fig:lifetime}.


\subsubsection{Bias}

Using the same procedure, a lifetime fit has been performed on a signal
Monte Carlo sample, giving $\tau_{B^0} = 1.48 \pm 0.025$\ps.
Removing the tracks from the $D^0$, the bias is removed and the result,
 $\tau_{B^0} = 1.52 \pm 0.02$\ps, is compatible with the generated
lifetime $\tau_{B^0} = 1.56$\ps. The estimated bias is then 0.04\ps\ 
and this has been added to the fit result from data to obtain the quoted final 
lifetime value.

Table~\ref{table:conecut} shows the fit results obtained with different values for
the cone cut, after applying the appropriate
Monte Carlo correction for bias. The corrected lifetime is stable with 
respect to the variation of the cone cut. 

\begin{table} [t]
\caption{Results of the fit obtained with different values for 
the cone cut for a smaller data sample corresponding to 5.3 \invfb\, after
applying the Monte Carlo bias correction. All the times are in ps.} \vspace{0.3cm}
  \centering
  \begin{tabular}{|l|c|c|c|c|}
    \hline
    Cut (rad)  & Events &  $\tau_{raw} $ & $\Delta \tau_{MC}  $&  $\tau_{corr} $ \\
    \hline\hline
1 &  2595 &  $1.42 \pm 0.04  $ & $0.12 \pm 0.02  $ & 1.54  \\ 
1.5 &  1839 &  $1.49\pm 0.05 $ & $0.06 \pm 0.02 $ &1.55  \\ 
2 &  1287 &  $1.52\pm 0.06  $ & $0.04 \pm 0.02 $ &1.56  \\ 
    \hline
  \end{tabular}
\label{table:conecut}
\end{table}

\subsubsection{Systematics}
A preliminary evaluation of the systematic errors on the lifetime result
has been performed. 

The combinatorial background fraction $\alpha_{back}$ has been varied
by 6\% leading to a variation of 0.039\ps\ in the fitted lifetime. 
The parametrisation of the combinatorial background  has been varied yielding variations of 0.012 ($\alpha_B$ \BB\ fraction), 0.006 
(  $\alpha_{uds}$ uds fraction)
and  0.010\ps\ ($\tau_{back}$ background lifetime).


This analysis relies on the resolution function fit from the
Monte Carlo simulation and therefore is sensitive to possible
discrepancies between simulation and data. 
The global scale factor $S$  has been left free in the fit 
giving  $S=1.13 \pm 0.12$. 
The corresponding variation of the lifetime of 0.051\ps\ has been
added to the systematic error.

The total systematic uncertainty of 0.07\ps, is dominated 
by the uncertainty in the continuum background fraction and the
uncertainty in the vertex resolution function 
(table \ref{table:sys}).
As these two contributions are related to variations estimated on the data themselves,
further increases in the size of the data sample should reduce these
systematic errors.

\begin{table} [t]
\caption{Sources of the systematic error and their impact on $\tau$.} \vspace{0.3cm}
  \centering
  \begin{tabular}{|l|c|c|}
    \hline
    Source & Variation & $\Delta \tau$ (ps)  \\
    \hline\hline
 $\alpha_{back}$ & $27\pm6 \%$ & 0.039 \\
$\alpha_B$ & $49 \pm 3 \% $  & 0.012 \\
$\alpha_{uds}$ & $ 37 \pm 20 \% $  & 0.006 \\
$\tau_{back}$  & $1.55 \pm 0.05~ ps$   & 0.010 \\ 
$S$   & $1.00 \pm 0.13 $  & 0.051 \\
bias & $0.04 \pm 0.02 $ & 0.020  \\ 
    \hline
Total & - & 0.07 \\
    \hline
  \end{tabular}
\label{table:sys}
\end{table}

As a further cross-check of the resolution in $\Delta z$, 
a data-Monte Carlo comparison has
been performed for the sideband 
and for the off resonance data. The agreement is good in both
cases. 



%
%
%

\section{\boldmath \Bz\ to $\Dstar\,\ell\,\bar{\nu_\ell}$ analysis and results}



\subsection{Event and track selection and sample composition}
\label{sec:lds_Selection}
\subsubsection{\boldmath \BB\ event selection}
\label{s:lds_evtsel}

Hadronic events were selected by requiring that at least four charged tracks be present. 
To reduce the contamination from light quark and \ccbar\ events, 
the second Fox-Wolfram moment $R_2$ is required to be smaller than 0.5.

\subsubsection{Track selection}

Only charged particles tracks were employed for this analysis. 
High momentum tracks were reconstructed by matching hits in the Silicon Vertex Tracker (SVT) 
with track elements in the Drift Chamber (DCH). Low momentum tracks 
do not hit enough wires in the DCH due to the bending induced by the magnetic field. 
They are however reconstructed by the SVT stand-alone pattern recognition, 
with more than 50\% efficiency for momentum $\plab>70$\mevc.

\subsubsection{Particle identification}

Charged tracks are identified as electrons using information from the electromagnetic calorimeter 
(ratio of the energy released to the associated track momentum, transverse profile of the 
electromagnetic shower), from the energy loss in the DCH and from the Cerenkov 
angle as measured with the Detector of Internally Reflected Cherenkov light (DIRC). 
The efficiency for electron identification in the acceptance of the electromagnetic calorimeter
is about 90\%, with a hadron mis-identification probability $< 1$\%. 

Muons were selected by requiring penetration through most of the Instrumented Flux Return (IFR). 
The selection requirements 
result in $\sim 75\%$ efficiency for muon identification with 
$\sim 2\%$ hadron mis-identification probability. The Cherenkov light information in the 
DIRC was then employed to further reject mis-identified kaons, by requiring the likelihood
for the kaon hypothesis to be less than five percent.

Kaons are used for \Bz\ flavour tagging.
Among all charged particles with momentum $\plab > 500$\mevc, 
kaon candidates are selected on the basis of the energy loss in the DCH and in the SVT, as well as
observed Cherenkov angle and number of photons in the DIRC. True kaons were selected with 
$\sim 80$ \% efficiency, while $\sim 1 \%$ of the pions were mis-identified as kaons.
 
\subsubsection{\boldmath \BtoDs\ sample selection}
\label{s:lds_selection}
Only the charged pion and the lepton are required to reconstruct \BtoDs\ decays. The lepton candidate
is required to have momentum in the \FourS\ rest frame (\ks)
in the range $1.4 <\ks< 2.3$\gevc, while the \pstar had to satisfy the requirement  
$\ks < 190$\mevc.

Events were then selected by exploiting the kinematics of the decay. 
The \Bz\ four momentum is computed by assuming that the \Bz\ is produced at rest in the \yfs\ decay, thereby
neglecting the small boost of the \B\ meson in this frame, where its momentum is about 300\mevc. 
Due to the limited phase space available in the decay \DsptoDz, the \pstar is emitted inside a 
fairly restricted cone centered about the \Dsp direction in the laboratory frame. 
The \Dsp four-momentum can therefore be computed by approximating its polar and azimuth angles 
with those for the \pstar, and parametrising its momentum as a linear function of the \pstar momentum:
\be
\nonumber p(\Dsp) = \alpha + \beta \cdot p(\pstar)
\eeq
The \Dsp\ direction is determined with an accuracy of about 15\degrees, 
the error on the \Dsp\ momentum is about 400\mevc.
The neutrino escapes detection, but its invariant mass can be computed from the \Bz,\Dsp and $\ellp$
four-momenta from the relation:
\be
\nonumber \mnusq = ({\cal P}(\Bz) - {\cal P}(\Dsp) -{\cal P}(\ellp))^2
\eeq
Neglecting resolution effects, this quantity must be zero, while background events are spread 
over a wide range of \mnusq.

The \Bz\ decay point is determined by intersecting the \pstar and $\ellp$ tracks, 
constrained to the beam-spot position, which was determined on a run-by-run basis using Bhabha events.
Only events for which the probability of the vertex fit was greater than 1\% were further 
considered. This cut rejected about two thirds of the combinatorial background, 
while still retaining $\sim$ 80\% of the signal.

\subsubsection{Sample composition}
\label{s:lds_compo}

The combinatorial background is determined from the data using those events 
for which the $\ell$ and \pstar have same electric charge (``wrong-charge''). 
The contribution of the events from any decay process of the kind \mbox{$B\ra D^{*-} \ell^+ \nu X$}
(``resonant events'') in the mass band ($\mnusq > -2\,\mathrm{\GeVc}$) was determined 
by subtracting the wrong-charge from the right-charge sample, 
normalized in the  sideband region \mbox{$-10 <\mnusq< -5\,\mathrm{\GeVc}$}. A small correction 
($\sim$ 5\%) was applied to the wrong-charge on-peak events to account for events of the kind
$\Bz \rar \ell$, $\Bzb \rar \Dstar$, which populates the right- and
 wrong-charge samples differently.
The number of events obtained in the \FourS\ resonance data sample is reported in the first line of 
Table~\ref{t:lds_sample}. The second line shows the number of events obtained in the continuum data sample, 
multiplied by the ratio of the resonant and continuum luminosities: non \bb events account for
about 15\% of the combinatorial background in the mass band and 5\% of the sample obtained after 
subtraction of the combinatorial.

\begin{table}[!htb]
\caption{Data statistics. The number of continuum events is normalized by the ratio of 
integrated luminosities.} \vspace{0.3cm}
\begin{center}
\begin{tabular}{|l|c|c|c|} \hline
Data Sample & Total  & Resonant \\ \hline\hline
On-peak     & 190080 & 89360  \\
Off-Peak    &  23900 &  4160  \\ \hline
\end{tabular}
\end{center}
\label{t:lds_sample}
\end{table}

\begin{figure}[!htb]
\begin{center}\begin{tabular}{cc}
\includegraphics[height=7cm]{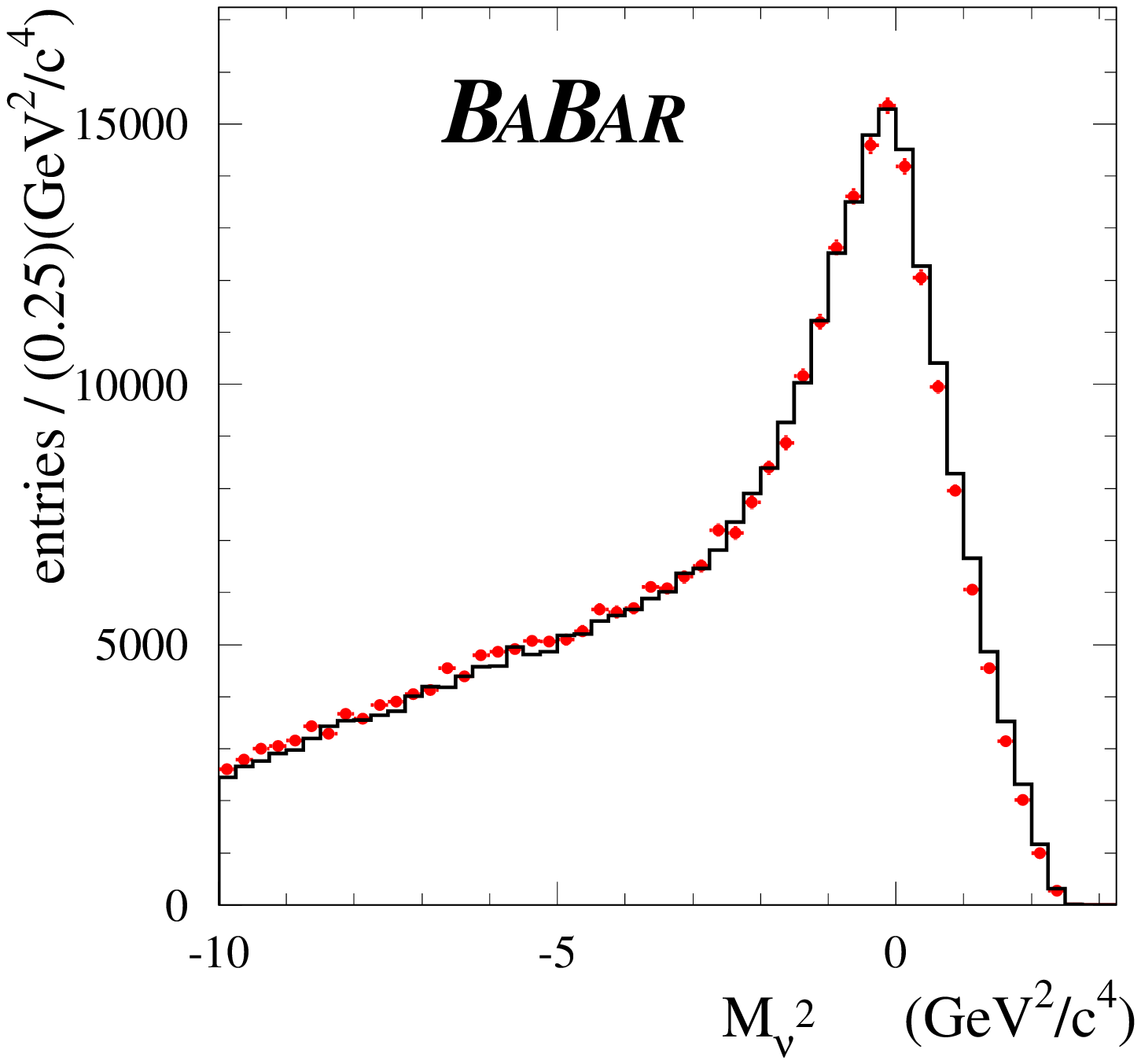} &
\includegraphics[height=7cm]{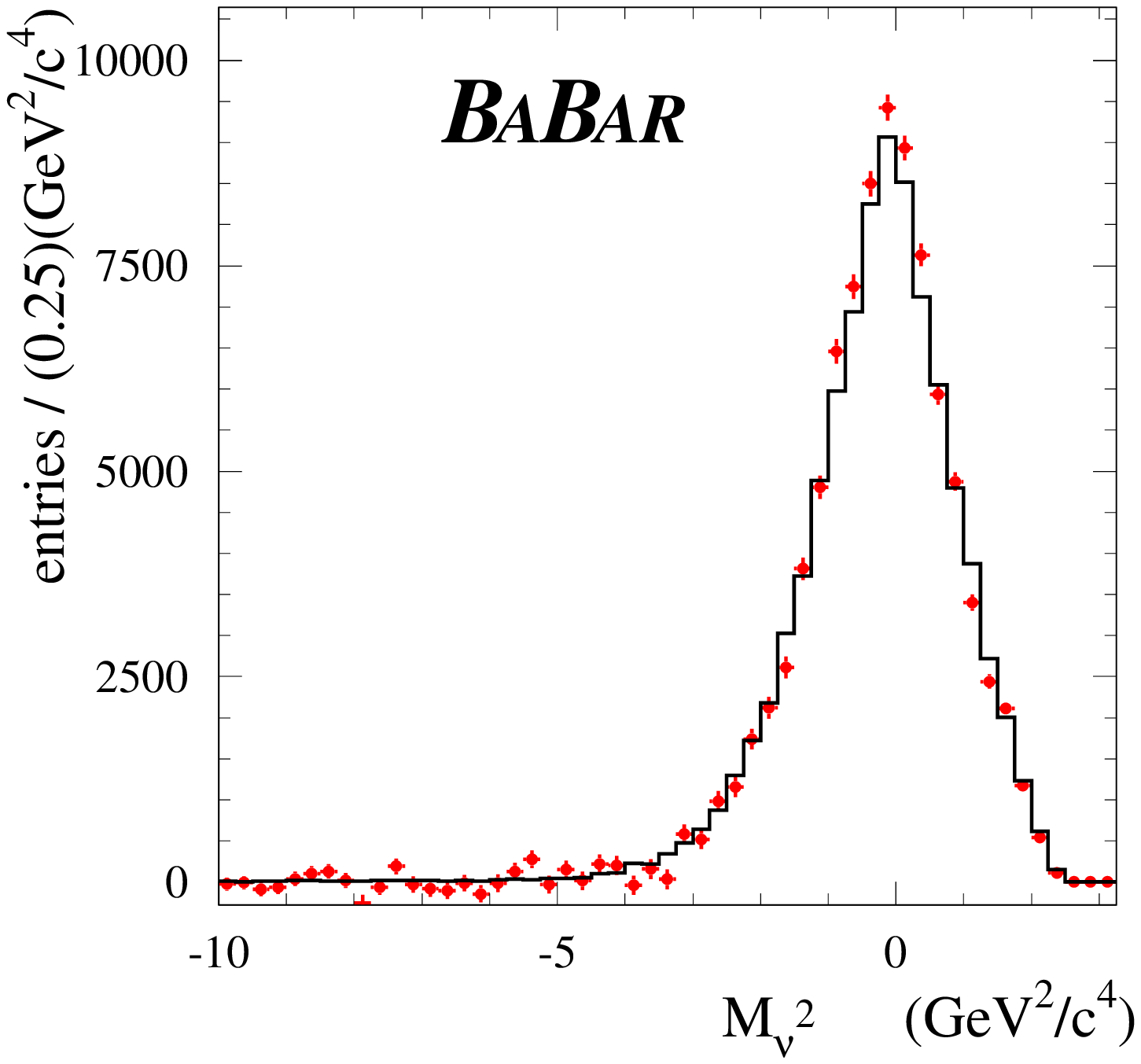}
\end{tabular} 
\caption{\mnusq\ distribution for the right charge data, after subtraction of 
the off-peak events for electron candidates. Dots with error bar: data; histogram: Monte Carlo
simulation. The right hand side plot is obtained after subtraction of the combinatorial
background.}
\label{f:lds_mm2}
\end{center}
\end{figure}

In the Monte Carlo simulation the
difference between the right charge and the rescaled wrong charge
events in the mass band is $-2.1 \pm 1.6$\%
No correction is applied to the data for this effect, and a relative systematic error 
of $\pm 2 \%$ is assumed. The number of resonant events in the sample is therefore:
\be
\N_{res} = 89360 \pm 500 ~\mathrm{stat.} \pm 1800 ~\mathrm{syst.}
\eeq
Figure \ref{f:lds_mm2} shows the \mnusq\ distribution for the electron data sample
after subtraction of the off-peak data, before (left) and after (right) 
subtracting the combinatorial background.

Several physical processes contribute to the $\Dstar\ell \nu$ final state. Contamination from
\Bp\ is mostly due to the decay $\Bp \rar \Dsst^0 \ell \nu$ where the orbital charm excited state,
$\Dsst^0$, decays into  $\Dsp \pi^+$. The fraction of these events is 9$\pm$5\%, where the error includes
the systematic uncertainty due to the efficiency of the \BtoDss\ selection. The composition of the 
sample is reported in Table~\ref{t:lds_compo}. All the processes originating from a
\Bz\ decay are considered as signal for this analysis.
\bt[!htb]
\caption[]{Sample Composition in the signal region (in \%).}
\bc \begin{tabular}{|l|c|} \hline 
$\BtoDs$                    & $36.6$ \\ 
$\Bzb\rar\Dsst$             & $ 2.7$ \\ 
$\Bzb\rar \Dsp X_c/\tau$    & $ 0.4$ \\ 
$B \rar \Dsp fake~l$        & $ 0.8$ \\ 
$\Bp\rar\Dsst$              & $ 4.5$ \\ 
off-peak resonant           & $ 2.4$ \\ \hline 
$D^*/l$ from different Bs   & $ 4.0$ \\ 
Other \bb\ Combinatorial    & $40.7$ \\
continuum                   & $ 7.9$ \\ \hline
\end{tabular} 
\label{t:lds_compo}
\ec\et

\subsection{\boldmath \tBz measurement and mixing}
\label{sec:lds_anal}

\subsubsection{Proper time determination}

The position of the \BtoDs\ ("decay") vertex is reconstructed as described above from the 
$\ell ,\pstar$ tracks and the beam spot. The decay point of the other \B\ ("tag" vertex)
is determined using the remaining tracks in the event, constrained to the decay vertex in
the plane orthogonal to the beam axis ($x,y$). This constraint exploits the fact that in the
$x$-$y$ view the separation between the \Bz\ and the \Bzb\ is small compared to the
experimental resolution; it is applied to remove the tracks from charm or hyperon 
decay from the vertex fit, which would bias the $\Delta z$ reconstruction. To reduce the additional bias, peculiar to 
inclusive analyses, due to the un-resolved particles from the \Dz\ produced at the decay vertex,
all the tracks lying within a one radian cone around the \pstar\ are rejected from the tag vertex fit.
In this way, about 60\% of the \Dz\ tracks are removed, while still retaining $\sim 70\%$ of
the \B\ tag tracks.

The \Bz\ lifetime is determined by measuring event by event the distance along $z$   
between the decay and the tag vertex, as already described in section 3.2.4.


In the Monte Carlo simulation the $\Delta z$ resolution for the signal events is described by
two Gaussian distributions: the narrow (wide) component containing about 65\% (35\%) 
of the events, with a width of 120 (370)\mum\ and a small bias of about 20 (65)\mum.
The pull distribution is parametrised by the sum of two Gaussian distributions, the narrow (wide) component 
contained 82\% (18\%) of the events, with a width of 1.07 (2.4) and a bias of 0.19 (0.7).
The same resolution was assumed for \Bp\ events.

\subsubsection{Determination of \boldmath \tBz}

The \Bz\ lifetime is determined by means of an unbinned maximum likelihood fit, accounting for the 
event-by-event error determined by the vertex reconstruction algorithm.
The fit function was the sum of the probability density function (pdf) for \Bz,\Bp\ and combinatorial
background multiplied by the relative fractions for each type in the sample.

The $\Delta z$ probability density function for \Bz\ events is described by the 
convolution of the proper time exponential with the pull resolution function described above.
The same function is assumed for \Bp\ events, with a lifetime fixed to the present world average
$\tau_\Bp =~ 1.65 \pm 0.04$\ps.

The combinatorial background is described by the sum 
of a zero-lifetime and a non-zero lifetime component, both convoluted with a double Gaussian resolution function.
All parameters (lifetime and fraction of non-zero lifetime events, widths, fractions and biases of 
the resolution function) were determined independently in the data and in the simulation 
by fitting wrong-charge events with the same program used to determine \tBz. Good consistency is
found in the data and in the simulation, except for the rms width of the resolution function, which 
is 1.06 times larger in data. This discrepancy is attributed to resolution effects.

The method was first tested on the dedicated sample of simulated \BtoDs\ events after applying the selection requirements 
described above. Due to the influence of the residual tracks from the un-reconstructed \Dz, the
measured lifetime in Monte Carlo is smaller than the generated one.
The resulting reconstruction bias $\DTau\ = 0.217 \pm 0.010$\ps\ for the lifetime is 
defined as the difference between the generated value of 1.560\ps\ and the fitted 
value $\tBz^{raw,MC} = 1.343 \pm 0.010$\ps.

\begin{figure}[!htb]
\begin{center}\begin{tabular}{cc}
\includegraphics[height=7cm]{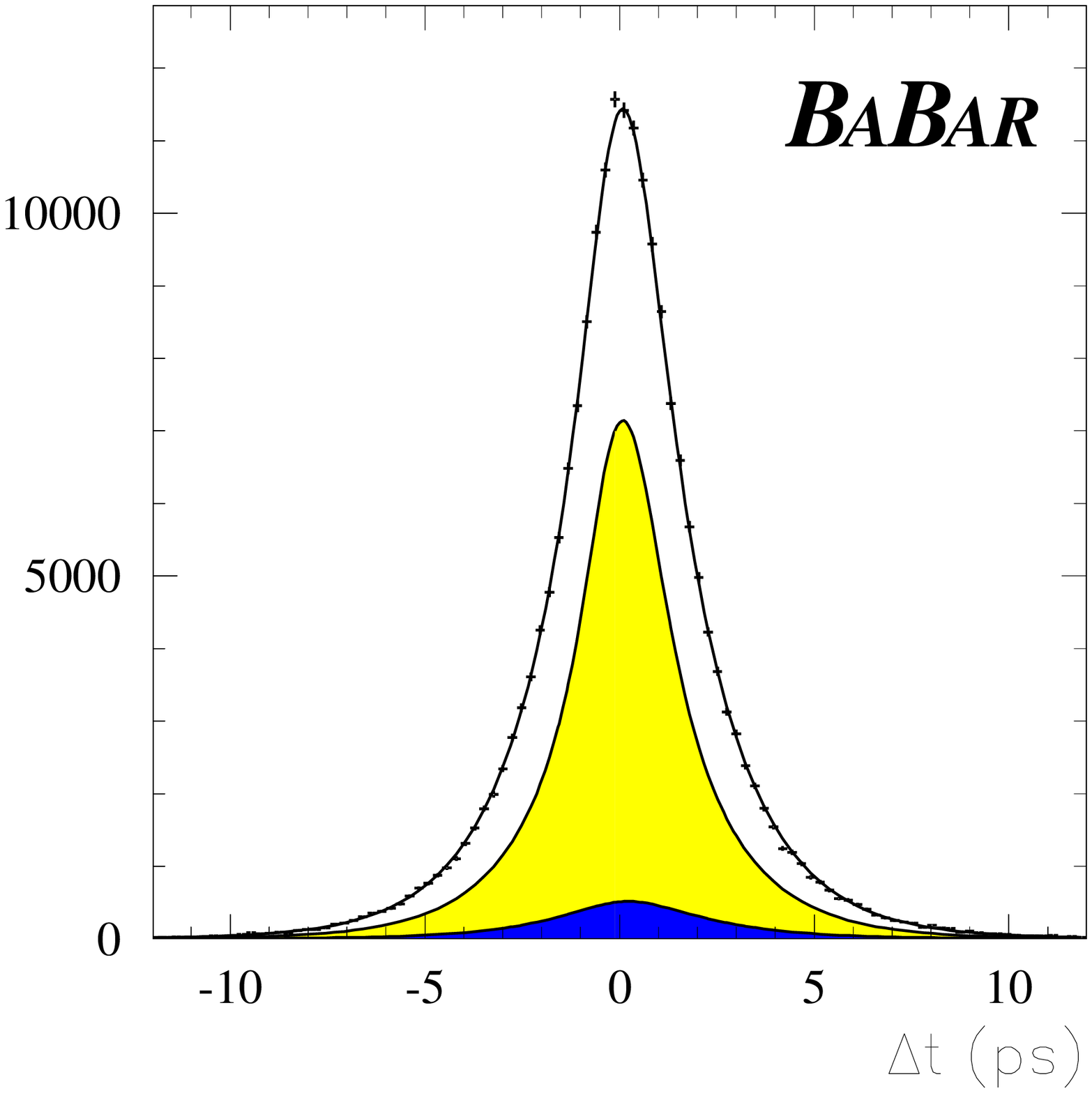}&
\includegraphics[height=7cm]{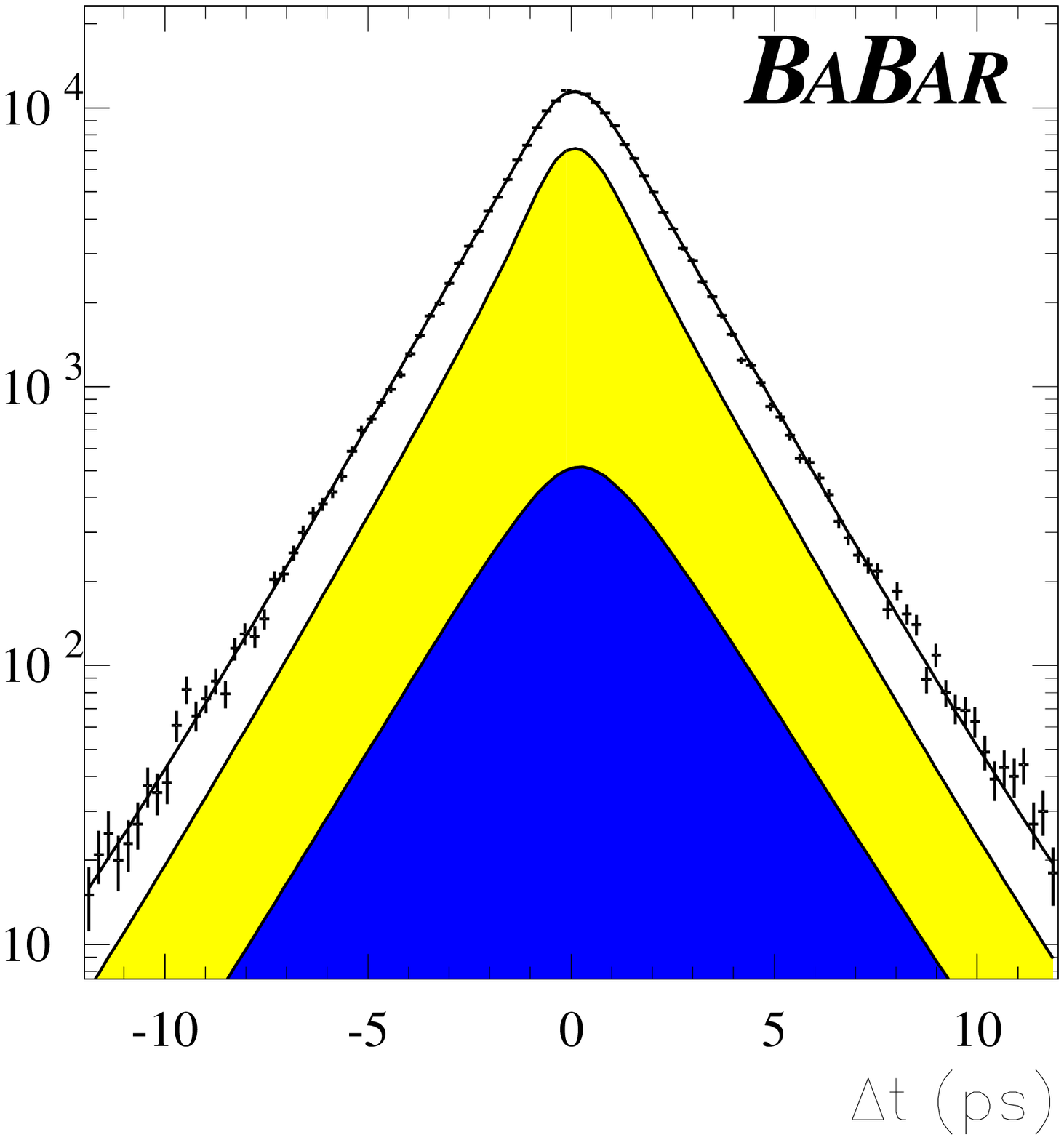}
\end{tabular}
\caption{ Lifetime fit result on a linear (left) and a logarithmic scale (right).
The inner dark area is the \Bp\ contribution, the lighter area is the combinatorial background,
the line is the result of the fit. Dots with error bars represent the data.} 
\label{f:lds_fitres}\end{center}\end{figure}

The final lifetime result for the data \tBz\ was obtained by
adding the bias \DTau\ determined previously in the Monte Carlo to the raw fit result 
$\tBz^{raw,Dt} = 1.405 \pm 0.016$\ps.
The lifetime obtained in this way is:
\be
\tBz = 1.622 \pm 0.016 ~\ps
\eeq
where the error is statistical only. Figure~\ref{f:lds_fitres} shows a comparison between the 
data and the fit result.

\subsection{Systematic studies}
\label{sec:lds_Systematics}
The study of the systematic effects is still preliminary. The following sources
of error have been considered:
\bi
\item {\bf Sample Composition.} The amount of \Bp\ is varied by $\pm 50 \%$;
the fraction of combinatorial background is varied by $\pm 2$ \% according to the study described
in Section \ref{sec:lds_Selection}; the fractions of 
events from uncorrelated $\ell$ \Dsp\ production falling in the right- and wrong-charge samples are
also varied within their allowed range, as computed from the measurements of
the branching ratios $\Br(B\rar\ell X)$, $\Br(B\rar\Dsp X)$, and of the \BzBzb\  mixing parameter $\chi_d$.  
\item {\bf Background Lifetime.} The \Bp\ lifetime is varied within its error;
the parameters of the pdf representing the combinatorial background is
varied within their allowed range, properly accounting for correlations: this gives an error
of $\pm0.02$ \ps.
Alternatively, the background parameters determined from a fit to the data side band are
used, and the difference in the result (+0.07 \ps ) is added in quadrature to the systematic error. 
\item{\bf Resolution Effects.} The fit was repeated by scaling the resolution function in the data by 
the factor 1.06 as indicated by the fit to the background sample: the measured lifetime decreased by
0.06\ps. This difference is considered as the systematic error due to the detector resolution. 
If the fraction of events in the narrow Gaussian describing the resolution function were varied by $\pm 5\%$
the fit result would vary by $\pm 0.04$\ps. This is currently taken as a rough estimate of error for the pdf in data.
The stability of the result versus some relevant variables ($\theta_{\ell,\ps},\phi_{\ell,\ps},
\plab_{\ell,\ps},\mnusq $) was also studied, but no significant effect was found.

\item{\bf Analysis Bias.} The sizeable correction for bias is determined from the Monte Carlo simulation. 
In the data, the fit was repeated for different values of the \pstar cone cut.
The systematic error was computed as half the difference between the largest and the 
smallest values obtained.\par
\ei
Table \ref{t:lds_syst} shows the components of the estimated systematic error.

\bt[!hbt]
\caption{Contributions to the estimated systematic error.}
\bc\begin{tabular}{|l|c|c|}  \hline
Source        & Variation     & $\sigma(\tBz)$ ps \\ \hline\hline
\Bp fraction  & $\pm 50 \%$   & $\mp 0.005$       \\
$B\rar\Dsp,\bar{B}\rar \ell$ &   & $\pm 0.002$  \\
Comb. frac.   & $\pm 2 \% $   & $\pm 0.014$     \\
Background pdf  &               & $^{+0.070}_{-0.020}$     \\
$\tau(\Bp)$   & $\pm 0.04 $ps & $\pm 0.005$     \\
Error Scale   &  1.06         & $ -0.060  $     \\
Data pdf      & $\pm 5\%$     & $ \pm 0.040$     \\
\DTau         &               & $ \pm 0.033$    \\ \hline
Total         &               & \rule[-1pt]{0mm}{14pt}$^{+0.091} _{-0.085}$ \\ \hline
\end{tabular}
\label{t:lds_syst}
\ec\et


\section{Summary}
A large sample of \BDstarpi\ and \BtoDs\ decays has been obtained from  
the data sample collected by the \babar\ detector at the \pep2\ 
storage ring by means of partial \Dsp\ reconstruction.  Only 
the low energy $\pi^-$ from the decay $D^{*-} \rightarrow \pi^- D^0$
was tagged and coupled to a high momentum opposite charge 
particle, either a pion or a lepton.

In the \BDstarpi\ case, the missing mass recoiling against the system of the 
two pions was determined using the beam energy constraint. A clear excess of 
$10922 \pm 235$ events in the $D^0$ mass range was observed, and 
interpreted as the signal of \BDstarpi\ production.

Events from the \BtoDs\ three body decay were selected by 
reconstructing the mass of the un-observed neutrino: for this purpose, the 
\Bz\ four momentum was computed neglecting its small boost from the
\FourS\ decay, and the \Dsp four momentum was computed by properly 
parametrising the one of the soft pion. A signal of 
$(89.4 \pm 0.5 \mathrm{stat.} \pm 1.8 \mathrm{syst.} ) \times 10^3$
events was observed.

The lifetime of the \Bz\ meson was determined from the distance, projected 
along the beam direction, between the tagged and the recoil \Bz\ meson 
in the event. The preliminary results are:
\ba
\nonumber \tBz &=& 1.55 \pm 0.05 \pm 0.07 \ps \quad (D^*\pi), \\
\nonumber \tBz &=& 1.62 \pm 0.02 \pm 0.09 \ps \quad (D^*\ell\nu),
\ea
were obtained by the two analyses. The two values are in good agreement,
while the mutual correlation is small.

Preliminary studies indicate sufficient tagging efficiency and flavor 
separation for a time dependent measurement of the \Bz\ mixing process. 
Clear evidence of a time dependent asymmetry is observed in the data, 
consistent with the expectation from the simulation. This will allow 
a precise determination of the \BzBzb\ mass difference \dmd\ in the 
future.


\section*{Acknowledgments}
\label{sec:Acknowledgments}

We are grateful for the contributions of our \pep2\ colleagues in
achieving the excellent luminosity and machine conditions
that have made this work possible.
We acknowledge support from the
Natural Sciences and Engineering Research Council (Canada),
Institute of High Energy Physics (China),
Commissariat \`a l'Energie Atomique and
Institut National de Physique Nucl\'eaire et de Physique des Particules
(France),
Bundesministerium f\"ur Bildung und Forschung
(Germany),
Istituto Nazionale di Fisica Nucleare (Italy),
The Research Council of Norway,
Ministry of Science and Technology of the Russian Federation,
Particle Physics and Astronomy Research Council (United Kingdom), the
Department of Energy (US),
and the National Science Foundation (US). In addition, individual support 
has been received from the Swiss 
National Foundation, the A. P. Sloan Foundation, the Research Corporation,
and the Alexander von Humboldt Foundation.
The visiting groups wish to thank 
SLAC for the support and kind hospitality
extended to them.

\end{document}